\errorstopmode
\input amssym.def
\input amssym.tex


\magnification=\magstephalf
\hsize=14.0 true cm
\vsize=19 true cm
\hoffset=1.0 true cm
\voffset=2.0 true cm

\abovedisplayskip=12pt plus 3pt minus 3pt
\belowdisplayskip=12pt plus 3pt minus 3pt
\parindent=1.0em


\font\sixrm=cmr6
\font\eightrm=cmr8
\font\ninerm=cmr9

\font\sixi=cmmi6
\font\eighti=cmmi8
\font\ninei=cmmi9

\font\sixsy=cmsy6
\font\eightsy=cmsy8
\font\ninesy=cmsy9

\font\sixbf=cmbx6
\font\eightbf=cmbx8
\font\ninebf=cmbx9

\font\eightit=cmti8
\font\nineit=cmti9

\font\eightsl=cmsl8
\font\ninesl=cmsl9

\font\sixss=cmss8 at 8 true pt
\font\sevenss=cmss9 at 9 true pt
\font\eightss=cmss8
\font\niness=cmss9
\font\tenss=cmss10

\font\eighttt=cmtt8

\font\sixmib=cmmib6
\font\sevenmib=cmmib7
\font\eightmib=cmmib8
\font\ninemib=cmmib9
\font\tenmib=cmmib10

 at 12 true pt
 at 12 true pt
\font\bigrm=cmr10 at 12 true pt
 at 12 true pt
 at 12 true pt

 at 16 true pt
 at 16 true pt
\font\Bigrm=cmr12 at 16 true pt
 at 16 true pt
 at 16 true pt

\catcode`@=11
\newfam\ssfam
\newfam\mibfam

\def\tenpoint{\def\rm{\fam0\tenrm}%
    \textfont0=\tenrm \scriptfont0=\sevenrm \scriptscriptfont0=\fiverm
    \textfont1=\teni  \scriptfont1=\seveni  \scriptscriptfont1=\fivei
    \textfont2=\tensy \scriptfont2=\sevensy \scriptscriptfont2=\fivesy
    \textfont3=\tenex \scriptfont3=\tenex   \scriptscriptfont3=\tenex
    \textfont\itfam=\tenit                  \def\it{\fam\itfam\tenit}%
    \textfont\slfam=\tensl                  \def\sl{\fam\slfam\tensl}%
    \textfont\bffam=\tenbf \scriptfont\bffam=\sevenbf
                           \scriptscriptfont\bffam=\fivebf
                           \def\bf{\fam\bffam\tenbf}%
    \textfont\ssfam=\tenss \scriptfont\ssfam=\sevenss
                           \scriptscriptfont\ssfam=\sevenss
                           \def\ss{\fam\ssfam\tenss}%
    \textfont\mibfam=\tenmib \scriptfont\mibfam=\sevenmib
                             \scriptscriptfont\mibfam=\sevenmib
                             \def\mib{\fam\mibfam\tenmib}%
    \normalbaselineskip=13pt
    \setbox\strutbox=\hbox{\vrule height8.5pt depth3.5pt width0pt}%
    \let\big=\tenbig
    \normalbaselines\rm}

\def\ninepoint{\def\rm{\fam0\ninerm}%
    \textfont0=\ninerm      \scriptfont0=\sixrm
                            \scriptscriptfont0=\fiverm
    \textfont1=\ninei       \scriptfont1=\sixi
                            \scriptscriptfont1=\fivei
    \textfont2=\ninesy      \scriptfont2=\sixsy
                            \scriptscriptfont2=\fivesy
    \textfont3=\tenex       \scriptfont3=\tenex
                            \scriptscriptfont3=\tenex
    \textfont\itfam=\nineit \def\it{\fam\itfam\nineit}%
    \textfont\slfam=\ninesl \def\sl{\fam\slfam\ninesl}%
    \textfont\bffam=\ninebf \scriptfont\bffam=\sixbf
                            \scriptscriptfont\bffam=\fivebf
                            \def\bf{\fam\bffam\ninebf}%
    \textfont\ssfam=\niness \scriptfont\ssfam=\sixss
                            \scriptscriptfont\ssfam=\sixss
                            \def\ss{\fam\ssfam\niness}%
    \textfont\mibfam=\ninemib \scriptfont\mibfam=\sixmib
                            \scriptscriptfont\mibfam=\sixmib
                            \def\mib{\fam\mibfam\ninemib}%
    \normalbaselineskip=12pt
    \setbox\strutbox=\hbox{\vrule height8.0pt depth3.0pt width0pt}%
    \let\big=\ninebig
    \normalbaselines\rm}

\def\eightpoint{\def\rm{\fam0\eightrm}%
    \textfont0=\eightrm      \scriptfont0=\sixrm
                             \scriptscriptfont0=\fiverm
    \textfont1=\eighti       \scriptfont1=\sixi
                             \scriptscriptfont1=\fivei
    \textfont2=\eightsy      \scriptfont2=\sixsy
                             \scriptscriptfont2=\fivesy
    \textfont3=\tenex        \scriptfont3=\tenex
                             \scriptscriptfont3=\tenex
    \textfont\itfam=\eightit \def\it{\fam\itfam\eightit}%
    \textfont\slfam=\eightsl \def\sl{\fam\slfam\eightsl}%
    \textfont\bffam=\eightbf \scriptfont\bffam=\sixbf
                             \scriptscriptfont\bffam=\fivebf
                             \def\bf{\fam\bffam\eightbf}%
    \textfont\ssfam=\eightss \scriptfont\ssfam=\sixss
                             \scriptscriptfont\ssfam=\sixss
                             \def\ss{\fam\ssfam\eightss}%
    \textfont\ttfam=\eighttt \def\tt{\fam\ttfam\eighttt}%
    \textfont\mibfam=\eightmib \scriptfont\mibfam=\sixmib
                             \scriptscriptfont\mibfam=\sixmib
                             \def\mib{\fam\mibfam\eightmib}%
    \normalbaselineskip=10pt
    \setbox\strutbox=\hbox{\vrule height7.0pt depth2.0pt width0pt}%
    \let\big=\eightbig
    \normalbaselines\rm}

\def\tenbig#1{{\hbox{$\left#1\vbox to8.5pt{}\right.\n@space$}}}
\def\ninebig#1{{\hbox{$\textfont0=\tenrm\textfont2=\tensy
                       \left#1\vbox to7.25pt{}\right.\n@space$}}}
\def\eightbig#1{{\hbox{$\textfont0=\ninerm\textfont2=\ninesy
                       \left#1\vbox to6.5pt{}\right.\n@space$}}}

\font\sectionfont=cmbx10
\font\subsectionfont=cmti10

\def\figurecaptionfont{\ninepoint}
\def\tablecaptionfont{\ninepoint}
\def\footnotefont{\eightpoint}


\newcount\equationno
\newcount\bibitemno
\newcount\figureno
\newcount\tableno

\equationno=0
\bibitemno=0
\figureno=0
\tableno=0


\footline={\ifnum\pageno=0{\hfil}\else
{\hss\rm\the\pageno\hss}\fi}


\def\section #1. #2 \par
{\vskip0pt plus .10\vsize\penalty-100 \vskip0pt plus-.10\vsize
\vskip 1.6 true cm plus 0.2 true cm minus 0.2 true cm
\global\def\equationlabel{#1}
\global\equationno=0
\leftline{\sectionfont #1. #2}\par
\immediate\write\terminal{Section #1. #2}
\vskip 0.7 true cm plus 0.1 true cm minus 0.1 true cm
\noindent}


\def\subsection #1 \par
{\vskip0pt plus 1.0 true cm\penalty-50 \vskip0pt plus-1.0 true cm
\vskip2.5ex plus 0.1ex minus 0.1ex
\leftline{\subsectionfont #1}\par
\immediate\write\terminal{Subsection #1}
\vskip1.0ex plus 0.1ex minus 0.1ex
\noindent}


\def\appendix #1. #2 \par
{\vskip0pt plus .10\vsize\penalty-100 \vskip0pt plus-.10\vsize
\vskip 1.6 true cm plus 0.2 true cm minus 0.2 true cm
\global\def\equationlabel{\hbox{\rm#1}}
\global\equationno=0
\leftline{\sectionfont Appendix #1. #2}\par
\immediate\write\terminal{Appendix #1. #2}
\vskip 0.7 true cm plus 0.1 true cm minus 0.1 true cm
\noindent}



\def\equation#1{$$\displaylines{\qquad #1}$$}
\def\enum{\global\advance\equationno by 1
\hfill\llap{{\rm(\equationlabel.\the\equationno)}}}
\def\noenum{\hfill}

\def\nexteq#1{\cr\noalign{\vskip#1}\qquad}


\def\ifundefined#1{\expandafter\ifx\csname#1\endcsname\relax}

\def\ref#1{\ifundefined{#1}?\immediate\write\terminal{unknown reference
on page \the\pageno}\else\csname#1\endcsname\fi}

\newwrite\terminal
\newwrite\bibitemlist

\def\bibitem#1#2\par{\global\advance\bibitemno by 1
\immediate\write\bibitemlist{\string\def
\expandafter\string\csname#1\endcsname
{\the\bibitemno}}
\item{[\the\bibitemno]}#2\par}

\def\beginbibliography{
\vskip0pt plus .15\vsize\penalty-100 \vskip0pt plus-.15\vsize
\vskip 1.2 true cm plus 0.2 true cm minus 0.2 true cm
\leftline{\sectionfont References}\par
\immediate\write\terminal{References}
\immediate\openout\bibitemlist=biblist
\frenchspacing\parindent=1.8em
\vskip 0.5 true cm plus 0.1 true cm minus 0.1 true cm}

\def\endbibliography{
\immediate\closeout\bibitemlist
\nonfrenchspacing\parindent=1.0em}


\def\figurecaption#1{\global\advance\figureno by 1
\narrower\figurecaptionfont Fig.~\the\figureno. #1}

\def\tablecaption#1{\global\advance\tableno by 1
\centerline{\tablecaptionfont Table~\the\tableno. #1}}

\def\thicktablerule{\hrule height0.8pt}
\def\thintablerule{\hrule height0.4pt}

\tenpoint

\immediate\openin\bibitemlist=biblist
\ifeof\bibitemlist\immediate\closein\bibitemlist
\else\immediate\closein\bibitemlist
\input biblist \fi


\def\thismonth{\ifcase\month\or
January\or February\or March\or April\or May\or June\or
July\or August\or September\or October\or November\or December\fi}

\input epsf
\epsfclipon



\def\rmd{{\rm d}}

\def\urltilde{\kern -.15em\lower .7ex\hbox{\~{}}\kern .04em}



\def\proof{\noindent{\sl Proof:}\kern0.6em}

\def\frac#1#2{\hbox{$#1\over#2$}}
\def\dual{\mathstrut^*\kern-0.1em}

\def\lvec#1{\setbox0=\hbox{$#1$}
    \setbox1=\hbox{$\scriptstyle\leftarrow$}
    #1\kern-\wd0\smash{
    \raise\ht0\hbox{$\raise1pt\hbox{$\scriptstyle\leftarrow$}$}}
    \kern-\wd1\kern\wd0}
\def\rvec#1{\setbox0=\hbox{$#1$}
    \setbox1=\hbox{$\scriptstyle\rightarrow$}
    #1\kern-\wd0\smash{
    \raise\ht0\hbox{$\raise1pt\hbox{$\scriptstyle\rightarrow$}$}}
    \kern-\wd1\kern\wd0}
\def\cvec#1{\kern-0.5pt\vec{\kern0.5pt #1}}

\def\slash#1{\setbox2=\hbox{$\displaystyle#1$}%
             \setbox3=\hbox{$\displaystyle/$}%
             #1\kern-0.8\wd2/\kern-1.0\wd3\kern0.8\wd2\kern0.5pt}

\def\wick#1{\setbox2=\hbox{$\displaystyle#1$}
    \setbox3=\null\ht3=3.0pt\dp3=0.0pt\wd3=20.0pt
    #1\kern-\wd2\kern3.0pt\raise11.0pt\vbox{\hrule height0.3pt
    \hbox{\vrule width0.3pt\box3\vrule width0.3pt}}\kern-24.0pt\kern\wd2}

\def\longwick#1{\setbox2=\hbox{$\displaystyle#1$}
    \setbox3=\null\ht3=3.0pt\dp3=0.0pt\wd3=27.0pt
    #1\kern-\wd2\kern3.0pt\raise11.0pt\vbox{\hrule height0.3pt
    \hbox{\vrule width0.3pt\box3\vrule width0.3pt}}\kern-31.0pt\kern\wd2}

\def\verylongwick#1{\setbox2=\hbox{$\displaystyle#1$}
    \setbox3=\null\ht3=3.0pt\dp3=0.0pt\wd3=43.0pt
    #1\kern-\wd2\kern3.0pt\raise11.0pt\vbox{\hrule height0.3pt
    \hbox{\vrule width0.3pt\box3\vrule width0.3pt}}\kern-47.0pt\kern\wd2}


\def\nabstar#1{{\nabla\kern0.5pt\smash{\raise 4.5pt\hbox{$\ast$}}
               \kern-5.5pt_{#1}}}

\def\drvstar#1{{\partial\kern0.5pt\smash{\raise 4.5pt\hbox{$\ast$}}
               \kern-6.0pt_{#1}}}
\def\sdrvstar#1{{\partial\kern0.4pt\smash{\raise 3.6pt\hbox{$\ast$}}
                \kern-4.8pt_{#1}}}

\def\ldrvstar#1{{\lvec{\,\partial}\kern-0.5pt\smash{\raise 4.5pt\hbox{$\ast$}}
               \kern-5.0pt_{#1}}}


\def\MSbar{\overline{\rm MS\kern-0.5pt}\kern0.5pt}



\def\psibar{\overline{\psi}}

\def\zetabar{\overline{\zeta}}


\def\diracstar#1#2{
    \setbox0=\hbox{$\gamma$}\setbox1=\hbox{$\gamma_{#1}$}
    \gamma_{#1}\kern-\wd1\kern\wd0
    \smash{\raise4.5pt\hbox{$\scriptstyle#2$}}}


\def\SUthree{{\rm SU(3)}}



\def\Dw{D_{\rm w}}
\def\Dwdag{{\Dw}\kern-4pt^{\dagger}\kern1pt}
\def\Dm{D}
\def\Dmdag{\Dm^{\dagger}\kern-1pt}


\def\dgm{{\cal D}}
\def\dgmr{{\cal R}}
\def\uset{L}
\def\rset{R}


\def\vxl{{\cal V}}
\def\lnl{{\cal L}}

\def\nv{n_v}
\def\nl{n_l}
\def\nb{n_b}
\def\nlp{n_F}
\def\sF#1{F_{#1}}
\def\sT#1{T_{#1}}


\def\Nf{N_{\rm f}}

\def\lbar{\bar{l}}
\def\Sn{{\cal S}_n}
\def\tmap{\phi}

%
\rightline{CERN-PH-TH-2014-239}
\vskip1.2cm
\centerline{\Bigrm Instantaneous stochastic perturbation theory}

\vskip 0.6 true cm
\centerline{\bigrm Martin L\"uscher$^{\rm a,b}$}

\vskip1.5ex
\centerline{{\it $^{\rm a}$CERN,
Physics Department, 1211 Geneva 23, Switzerland}}

\vskip1.0ex
\centerline{{\it $^{\rm b}$Albert Einstein Center for Fundamental Physics}}
\centerline{{\it Institute for Theoretical Physics,
Sidlerstrasse 5, 3012 Bern, Switzerland}}

\vskip 0.8 true cm
\thintablerule
\vskip 2.0ex
\ninepoint
\leftline{\bf Abstract}
\vskip 1.0ex\noindent
A form of stochastic perturbation theory is described, where the
representative stochastic fields are generated instantaneously
rather than through a Markov process.
The correctness of the procedure is established
to all orders of the expansion and for a wide class of field theories
that includes all common formulations of lattice QCD.

\vskip 2.0ex
\thintablerule

\tenpoint


\section 1. Introduction

Stochastic perturbation theory has its roots
in stochastic quantization [\ref{SQI},\ref{SQII}],
but is nowadays mainly used as a tool in numerical lattice field theory
[\ref{SPThI},\ref{SPThII}] (see [\ref{DiRenzoScorzato}] for a
review and [\ref{BrambillaBridaEtAl},\ref{BridaHesse}] for
recent applications of the method). As in stochastic quantization,
the starting point is the Langevin equation
and thus a Markov process that simulates the field theory
considered. The stochastic field is then expanded in
a formal power series in the couplings, which allows the equation
to be solved order by order in the interactions.

Numerical stochastic perturbation theory has long proved to
be very useful. Often higher orders in the expansion
can be reached than would be practically feasible with
other methods.
As the lattice spacing $a$ is taken to zero,
the required numerical effort is however rapidly increasing.
In four dimensions, and taking autocorrelations into account,
the cost of the calculations tends to grow approximately
like $a^{-6}$.
Moreover, even if a higher-order
scheme is used for the numerical integration of the Langevin equation
[\ref{LvinI}--\ref{LvinIII}],
the simulation results must eventually be extrapolated to vanishing
integration step size in order to be safe of uncontrolled
systematic effects
[\ref{LvinIV}].

The form of stochastic perturbation theory described in
the following sections is
unrelated to stochastic quantization and is instead based on
the concept of a ``trivializing map'' [\ref{TrivMap}].
Such maps transform trivial random fields into stochastic fields
representing an interacting field theory in the same way as the fields
generated in a simulation do. Field configurations obtained in this
way from uncorrelated random fields are statistically independent
and thus provide an instantaneous simulation of the theory.

Trivializing maps practically solve the
theory and cannot be expected to be easily constructible.
There are, however, explicit and fairly simple
maps that trivialize the theory to any finite order of
perturbation theory.
In the case of the one worked out in this paper,
the trivializing stochastic field is given by a series of
rooted tree diagrams with random fields
attached to their leaves. Such diagrams
can be computed through a recursive procedure, where one starts
from the leaves of the diagram and progresses from one vertex to the next
until the root line is reached.
The trivializing field thus becomes accessible to numerical
evaluation.

The principal goal in this paper is to
establish the existence of trivializing maps of this kind and
to provide the theoretical background for the automatic
generation of the tree diagrams and the computation of their
coefficients\kern1pt\footnote{$\dagger$}{\footnotefont%
A program package providing this functionality for lattice QCD
and other theories can be downloaded from
{\tt http://cern.ch/luscher/ISPT}.}.

\section 2. Scalar field theory to second order --- an illustrative example

The concepts and strategies underlying instantaneous stochastic perturbation
theory are best explained in the case of a one-component
scalar field $\varphi(x)$ with (Euclidean) action
\equation{
  S(\varphi)=
  \int\rmd^4x\,\left\{\frac{1}{2}\partial_{\mu}\varphi(x)\partial_{\mu}\varphi(x)
  +\frac{1}{2}m^2\varphi(x)^2+{g\over4!}\varphi(x)^4\right\},
  \enum
}
$g$ being the coupling constant and $m$ the mass parameter of the theory.
An ultraviolet regularization is assumed in the following
and the perturbation expansion should be the renormalized
rather than the bare one, but for simplicity
these complications are skipped over in this section.

\subsection 2.1 Trivializing maps

Let $\eta(x)$ be a Gaussian random field with
mean zero and variance
\equation{
  \langle\eta(x)\eta(y)\rangle=\delta(x-y).
  \enum
}
In its simplest form,
a trivializing map is a transformation of fields, where
the ran\-dom field $\eta(x)$ is transformed to a field $\tmap(x)$
in an invertible manner such that
\equation{
  \langle\tmap(x_1)\ldots\tmap(x_n)\rangle
  =\langle\varphi(x_1)\ldots\varphi(x_n)\rangle
  \enum
}
for all $n$ and $x_1,\ldots,x_n$. The correlation function on
the right of this equation is the one determined by the action
(2.1) and the associated functional integral, while on the left
the product $\tmap(x_1)\ldots\tmap(x_n)$
is to be averaged over the Gaussian random field.

In the functional integral,
trivializing maps of this kind may be considered to be
transformations of integration variables.
A sufficient condition for eq.~(2.3) to hold is then
\equation{
  S(\tmap)-\ln\det\{\delta\tmap/\delta\eta\}=\int\rmd^4x\,\frac{1}{2}\eta(x)^2
  +\hbox{constant},
  \enum
}
where $\det\{\delta\tmap/\delta\eta\}$ denotes the Jacobian of the
transformation.

A more general class of trivializing maps is obtained by allowing
the trivializing stochastic field to depend on more than one
random field. Since the mapping
goes from several random fields to one stochastic
field in this case, it cannot be considered to be a transformation
of integration variables, but eq.~(2.3)
remains meaningful if the product of fields on the left of the equation
is averaged over all random fields.

\subsection 2.2 Trivializing map to second order in the coupling

To any finite order in the coupling,
trivializing stochastic fields can be obtained
in the form of a power series
\equation{
  \tmap(x)=\sum_{k=0}^{\infty}
  \tmap_k(x),
  \qquad
  \tmap_k(x)\propto g^k,
  \enum
}
with leading-order term
\equation{
  \tmap_0(x)=(-\Delta+m^2)^{-1/2}\eta(x).
  \enum
}
In particular,
there exist trivializing maps, where the fields
$\tmap_k(x)$ are linear combinations of Feynman diagrams of the
type shown in fig.~1.
The first diagram, for example, is equal to
\equation{
  -g\int\rmd^4y\,G_0(x,y)\tmap_0(y)^3,
  \enum
}
where $G_0(x,y)$ denotes the free propagator in position
space. Similarly, the diagram number $5$ evaluates to
\equation{
  (-g)^2\int\rmd^4y\,\rmd^4z\,G_0(x,y)G_0(y,z)^2\tmap_0(y)
  \tmap_0(z)^2.
  \enum
}
No symmetry factors are associated with these diagrams and their
values are calculated simply by integrating the product of
the propagators, vertices and attached random fields over the positions
of the vertices.

\topinsert
\vbox{
\vskip0ex

\epsfxsize=12.0cm\hskip0.0cm\epsfbox{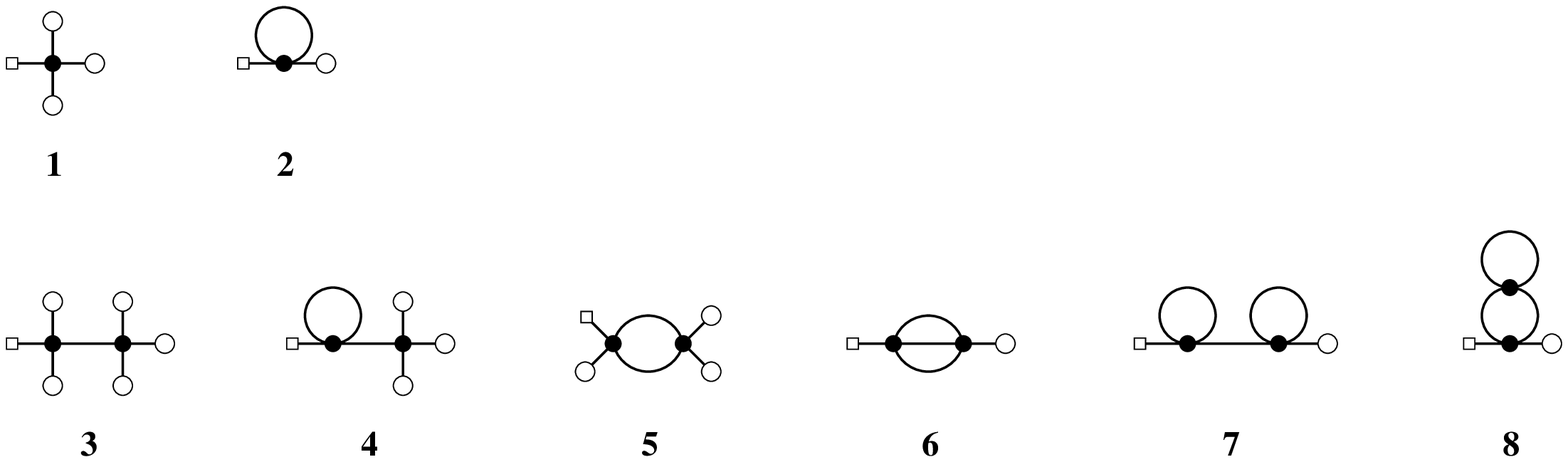}

\vskip2.5ex
\figurecaption{%
Diagrams $\dgm_1,\ldots,\dgm_8$
contributing to $\tmap_1(x)$ (diagrams 1 and 2) and $\tmap_2(x)$
(diagrams 3--8).
The values of these diagrams are determined by the standard Feynman rules
in the theory with action (2.1) except for
the external lines ending in open circles, which stand
for an insertion of the leading-order field (2.6).
A little square is attached to
the external lines that end at the point $x$.
}
}
\vskip0ex
\endinsert

The first- and second-order terms in the expansion (2.5) of
the trivializing map are then given by
\equation{
  \tmap_1(x)=\frac{1}{24}v(x;\dgm_1)+\frac{1}{8}v(x;\dgm_2),
  \enum
  \nexteq{3.0ex}
  \tmap_2(x)=
   \frac{7}{1152}v(x;\dgm_3)
  +\frac{1}{36}v(x;\dgm_4)
  +\frac{1}{96}v(x;\dgm_5)
  +\frac{1}{48}v(x;\dgm_6)
  \noenum
  \nexteq{2.0ex}
  {\phantom{\tmap_2(x)={}}}
  -\frac{1}{128}v(x;\dgm_7)
  +\frac{5}{64}v(x;\dgm_8),
  \enum
}
where $v(x;\dgm_1),\ldots,v(x;\dgm_8)$ are the values
of the diagrams drawn in fig.~1.
Insertion of these expressions in eq.~(2.4) shows that
the map does indeed trivialize the theory up to terms of
order $g^3$.
The correctness of the coefficients in eqs.~(2.9) and (2.10)
can also be checked by working out the two-point, four-point and six-point
correlation functions on both sides of eq.~(2.3) to second order in
the coupling.

\subsection 2.3 Breaking up the loops

The trivializing map described in the previous subsection is not
particularly useful, because its evaluation
at order $k$ in the coupling requires the computation
of Feynman diagrams with up to $k$ loops.
It is possible, however, to break up the loops by introducing
further Gaussian random fields
\equation{
  \eta_l(x),\quad l=1,2,\ldots,
  \enum
}
with mean zero and variance
\equation{
  \langle\eta_l(x)\eta_j(y)\rangle=\delta_{lj}\delta(x-y),
  \qquad
  \langle\eta(x)\eta_l(y)\rangle=0.
  \enum
}
The contraction of the labeled random fields at the leaves of
the tree diagrams shown in fig.~2 then reproduces the diagrams
in fig.~1.

\topinsert
\vbox{
\vskip0ex

\epsfxsize=12.0cm\hskip0.0cm\epsfbox{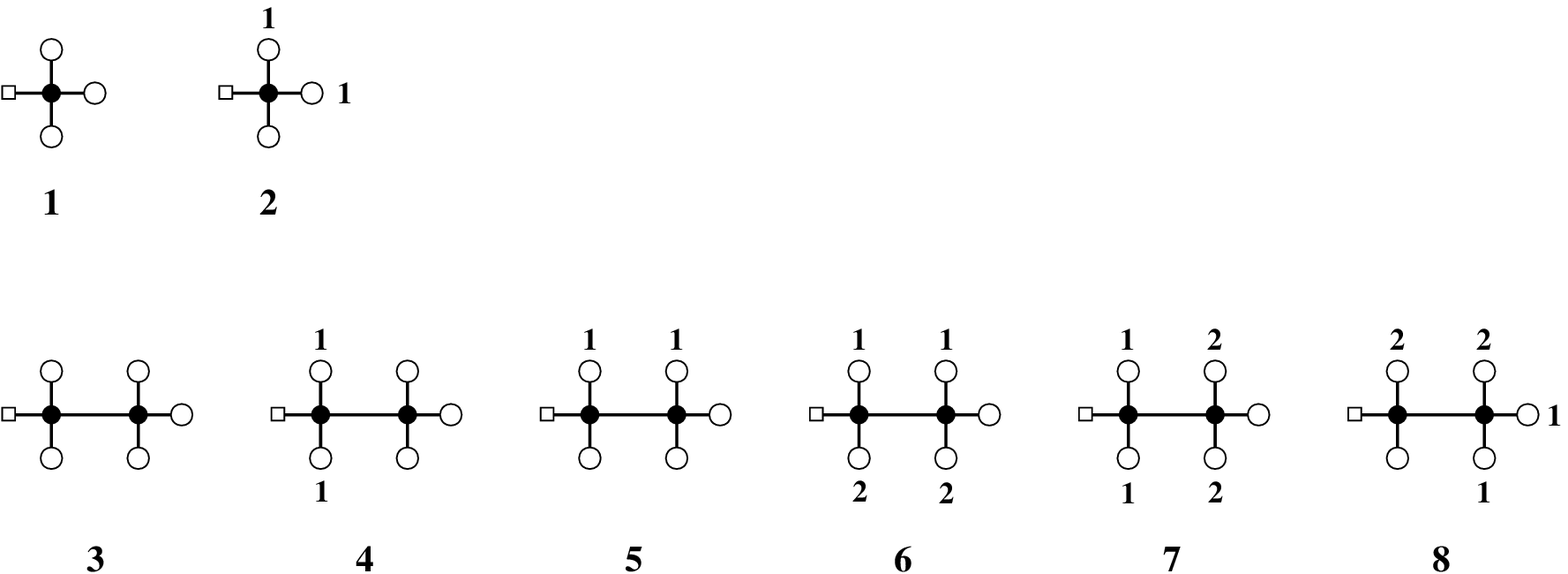}

\vskip2.5ex
\figurecaption{%
Labeled tree diagrams $\dgmr_1,\ldots,\dgmr_8$
corresponding to the diagrams shown in fig.~1.
Open circles with attached label $l$ represent the insertion of
the field (2.6) with the random field $\eta$ replaced by $\eta_l$.
Labeled open circles always
come in pairs, i.e.~each label occurs either zero or two times.
}
}
\vskip0ex
\endinsert

By replacing the diagrams $\dgm_1,\ldots,\dgm_8$
through the corresponding tree diagrams $\dgmr_1,\ldots,\dgmr_8$,
the trivializing field defined through eqs.~(2.6),(2.9) and (2.10)
becomes a stochastic field
that depends on $\eta$, $\eta_1$ and $\eta_2$.
One might expect this field to be trivializing too,
but a short calculation reveals that
eq.~(2.3) is violated at second order in the coupling,
because some contractions of the labeled random fields
generate terms that were previously not there.

As will be shown in the following sections,
the trivializing property of the stochastic field
can however be preserved by
adjusting the coefficients of the tree diagrams.
The diagram replacement and coefficient adjustment
amount to the substitutions
\equation{
  \tmap_1(x)\to\frac{1}{24}v(x;\dgmr_1)+\frac{1}{8}v(x;\dgmr_2),
  \enum
  \nexteq{3.0ex}
  \tmap_2(x)\to
   \frac{7}{1152}v(x;\dgmr_3)
  +\frac{1}{36}v(x;\dgmr_4)
  -\frac{1}{192}v(x;\dgmr_5)
  +\frac{7}{192}v(x;\dgmr_6)
  \noenum
  \nexteq{2.0ex}
  {\phantom{\tmap_2(x)\to{}}}
  -\frac{1}{128}v(x;\dgmr_7)
  +\frac{3}{32}v(x;\dgmr_8),
  \enum
}
and it is then again straightforward to check
that, after these substitutions, eq.~(2.3) still holds up to
higher-order terms%
\kern1pt\footnote{$\dagger$}{\footnotefont%
An interesting representation of the connected parts of the
correlation functions $\langle\varphi(x_1)\ldots\varphi(x_n)\rangle$
is obtained by replacing all connected Feynman diagrams
through the appropriate
labeled tree diagrams. Such non-factorized stochastic representations
are however
of limited use if the observables considered are complicated functions
of the fundamental fields.}.

\section 3. Perturbation expansion to all orders

Beyond the first few orders of perturbation theory,
the number of labeled tree diagrams and random-field contractions
rapidly becomes very large.
A mathematically precise terminology is then
required to be able to establish the existence
of trivializing maps
of the kind discussed at the end of sect.~2.

In this section, the class of field theories considered is
specified. Furthermore, a compact notation
is introduced that allows the Feynman rules to be stated
economically and in full generality.

\subsection 3.1 Abstract definition of the theory

Although instantaneous stochastic perturbation theory does not
require this, a lattice regularization
and a finite space-time volume are assumed from now on.
For simplicity of presentation, only the case of theories containing
both boson and fermion fields will be discussed.
Gauge symmetries (if any) should be fixed, with
ghost fields added as needed,
and the boundary conditions must be such that
the action has no zero modes at lowest order of perturbation theory.

The fields integrated over in the functional integral may carry
various indices.
It is convenient to treat the space-time coordinates
as further indices and to pack all indices into a
multi-index. There is then only one real boson field, $\varphi_a$,
depending on a multi-index $a$ that includes the
space-time coordinates,
the Lorentz and internal-symmetry indices and an
index labeling the different types boson fields.

Fermion fields (quark and ghost fields, in
particular) can be similarly packed into a single field $\psi_{\alpha}$
with multi-index $\alpha$. The associated antifermion field
$\psibar_{\beta}$ carries the same kind of multi-index and is
taken to be independent of the fermion field
(thus excluding Majorana fermions).

The total action $S$ of the theory is a function of the fields
$\varphi_a,\psi_{\alpha}$ and $\psibar_{\beta}$, which may depend on several
couplings\kern1pt\footnote{$\dagger$}{\footnotefont%
In theories with a non-trivial functional integration measure,
the action $S$ is assumed to include the terms that arise
from the expansion of the measure in powers of the coupling.}.
In perturbation theory, the coupling constants multiplying the
interaction terms are usually taken to zero
proportionally to one of them, say $g$, so that the
perturbation expansion of the action
\equation{
  S=\sum_{k=0}^{\infty}S_k,\qquad
  S_k\propto g^k,
  \enum
}
is an expansion in powers of $g$.
At this highly abstract level,
the bare and the re\-nor\-ma\-lized perturbation series only differ by
the parameterization of the various terms in the action.

\subsection 3.2 Propagators and vertices

To lowest order in the coupling, the action
\equation{
  S_0=\frac{1}{2}\varphi_aK^B_{ab}\varphi_b
  +\psibar_{\alpha}K^F_{\alpha\beta}\psi_{\beta}
  \enum
}
is assumed to be a non-degenerate quadratic expression in the fields
(here and below, the Einstein summation convention is
used for repeated field indices).
Moreover,
the kernel $K^B_{ab}$ must be a real symmetric matrix
with strictly positive eigenvalues.
The basic two-point functions
are then given by
\equation{
  \left.\langle \varphi_a\varphi_b\rangle\right|_{g=0}=D^B_{ab},
  \qquad
  K^B_{ab}D^B_{bc}=\delta_{ac},
  \enum
  \nexteq{2.0ex}
  \left.\langle \psi_{\alpha}\psibar_{\beta}\rangle\right|_{g=0}=
  D^F_{\alpha\beta},
  \qquad
  K^F_{\alpha\beta}D^F_{\beta\gamma}=\delta_{\alpha\gamma},
  \enum
}
at this order in the coupling.

The interaction terms in the action
are assumed to be of the general form
\equation{
  S_k=
  \sum_{j=2}^{k+2}{1\over j!}
  V^{(k,j)}_{a_1\ldots a_j}\varphi_{a_1}\ldots\varphi_{a_j}
  +\sum_{j=0}^k{1\over j!}
  W^{(k,j)}_{\alpha,\beta,a_1\ldots a_j}\psibar_{\alpha}\psi_{\beta}
  \varphi_{a_1}\ldots\varphi_{a_j},
  \enum
}
where the vertices $V^{(k,j)}_{a_1\ldots a_j}$ and
$W^{(k,j)}_{\alpha,\beta,a_1\ldots a_j}$ include the factor $g^k$
and are totally sym\-metric in the boson-field indices $a_1,\ldots,a_j$.
Four-fermion interactions are not explicitly included in the action,
but can be accommodated using Lagrange-multiplier fields.
Clearly, many components of the propagators and vertices may vanish
in a concrete case.

\topinsert
\vbox{
\vskip0ex

\epsfxsize=10.5cm\hskip0.75cm\epsfbox{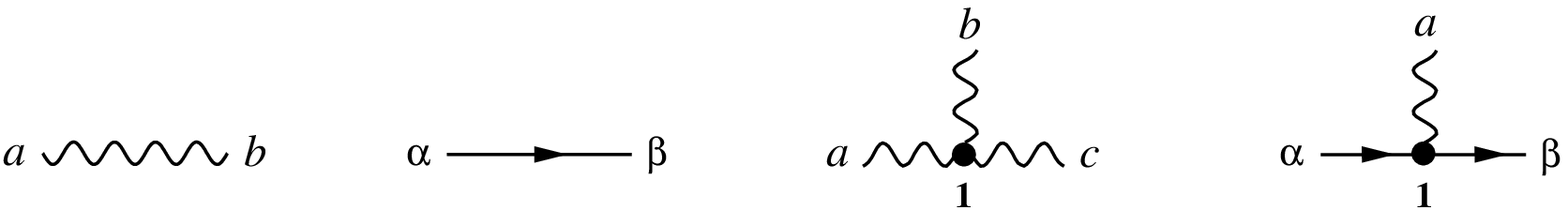}

\vskip1.5ex
\figurecaption{%
Graphical representation of the boson propagator $D^B_{ab}$, the
fermion propagator $D^F_{\alpha\beta}$, the boson vertex
$-V^{(1,3)}_{abc}$ and the fermion-boson vertex $-W^{(1,1)}_{\alpha\beta a}$.
An integer label is printed near the vertices
indicating their order in the coupling
(the label may be omitted if a vertex is
uniquely specified by the type and number of its legs).
}
}
\vskip1.0ex
\endinsert

\subsection 3.3 Feynman rules

The expansion of the basic correlation functions
\equation{
  \langle\varphi_{a_1}\ldots\varphi_{a_n}\psi_{\alpha_1}\ldots\psi_{\alpha_m}
  \psibar_{\beta_m}\ldots\psibar_{\beta_1}\rangle
  \enum
}
in Feynman diagrams may now be derived as usual from the functional
integral. The diagrams are built from the
propagators and vertices introduced above, with
their external lines labeled by the indices $a_1\ldots\beta_m$
of the fields in the correlation function (see figs.~3 and 4).
In general, the diagrams need not be connected and
there may be isolated lines,
i.e.~lines not attached to any vertex.

The value $v(\dgm)_{a_1\ldots\beta_m}$
of a diagram $\dgm$ is calculated by writing down the
expressions for each graphical element (lines and vertices)
and by contracting the indices of the vertices
with the appropriate ones of the propagators representing the
attached lines. At the other ends of external lines and the
ends of the isolated lines, the indices of the propagators
are set to the field indices $a_1,\ldots,\beta_m$ labeling the lines.

The contribution of the diagram to the correlation function (3.6)
is equal to the product of its value, $v(\dgm)_{a_1\ldots\beta_m}$,
the inverse of the symmetry factor $\sF{1}\sF{2}\sF{3}$
defined in sect.~4 and a fermion sign factor.
Each fermion loop contributes a factor $-1$ to the latter
and a further factor $-1$ is to be included if
the open fermion paths in the diagram associate the
fermion indices $\alpha_1,\ldots,\alpha_m$
with an odd permutation of the antifermion
indices $\beta_1,\ldots,\beta_m$ (as in the second diagram
in fig.~4).

It may be worth noting that the diagrams considered here
expand to possibly many ordinary
diagrams when passing to the notation commonly used in field theory.
In lattice QCD,
for example, the first diagram in fig.~4 includes two ordinary
diagrams, one with a quark and the other with a ghost loop, both
having the correct fermion-boson vertices.

\topinsert
\vbox{
\vskip0ex

\epsfxsize=8.0cm\hskip2.0cm\epsfbox{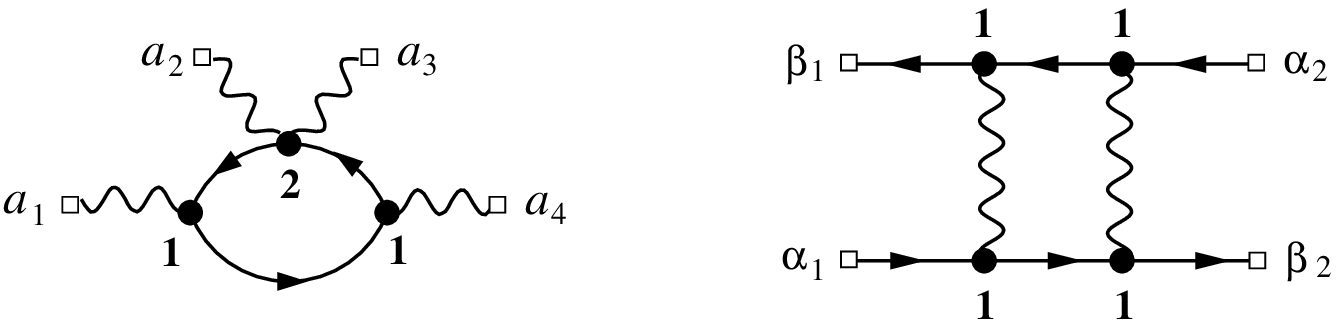}

\vskip2.0ex
\figurecaption{%
Two examples of diagrams complying with the rules stated in subsect.~3.3.
The squares at the ends of the external lines
indicate that the lines are not amputated.
}
}
\vskip0ex
\endinsert

\section 4. List description of Feynman diagrams

For the construction of trivializing stochastic fields
in the most general case,
a formal description of Feynman diagrams is helpful and
in any case required, if
the generation and structural analysis of the diagrams
is to be performed by a computer.

\subsection 4.1 Diagram data

Feynman diagrams may be described by
ordered lists of vertices and lines,
where the latter refer to the first in a way
explained below.

Vertices are distinguished
by their type (boson or fermion-boson), the order $k$
in the coupling and the number $j$ of boson legs.
An example of a vertex list is
\equation{
  \vxl=[V^{(1,3)},V^{(1,3)},V^{(2,2)},
  W^{(2,0)},W^{(2,4)}].
  \enum
}
Vertex lists merely contain the vertex symbols rather
than the vertices themselves.
No particular ordering is required and
a vertex symbol may appear more than once.

The list of lines may include internal, external and isolated lines
in any order. An example of a line list is
\equation{
   \lnl=[({\rm b},1,2),({\rm f},6,5),
   ({\rm f},5,\beta_1),({\rm f},\alpha_2,6),
   ({\rm b},a_1,a_2)],
   \enum
}
where the items $({\rm b},i_1,i_2)$ and $({\rm f},i_1,i_2)$ represent
boson and fermion lines, respectively. The indices $i_1,i_2$ can be
natural numbers or field indices.
A line is outgoing at the vertex number $i_1$
if $i_1$ is a natural number and ingoing at the vertex number $i_2$ if
$i_2$ is a natural number.
In the case of an external line, $i_1$ or $i_2$
is the field index labeling the line
(isolated lines have
both indices set to field indices).

The first diagram shown in fig.~4, for example, is described by the
lists
\equation{
  \vxl=[W^{(1,1)},W^{(1,1)},W^{(2,2)}],
  \noenum
  \nexteq{2.0ex}
  \lnl=[({\rm b},1,a_1),({\rm b},3,a_2),({\rm b},3,a_3),({\rm b},2,a_4),
        ({\rm f},1,2),({\rm f},2,3),({\rm f},3,1)].
  \enum
}
In general, for a vertex and line list to
consistently describe a diagram,
the number and type of lines attached to each vertex must
match the number and type of its
legs. Moreover, there may be no lines attached to inexistent
vertices.

\subsection 4.2 Vertex and line permutations

The description of a diagram through a pair $(\vxl,\lnl)$
of vertex and line lists implicitly goes along with a
labeling of its vertices and lines. In the following,
the numbers of vertices and lines will be
denoted by $\nv$ and $\nl$, respectively,
and the $i$'th vertex and $j$'th line are referred to as
$\vxl[i]$ and $\lnl[j]$.

There are two kinds of transformations of the diagram data
$(\vxl,\lnl)$, which leave the diagram unchanged:

\parindent=2.75em
\vskip1ex
\item{$(\sT{1})$} The vertices may be reordered in
an arbitrary way. This amounts to replacing the vertex list by
a new list $\vxl'$ such that
\equation{
  \vxl'[i]=\vxl[\sigma(i)]\hskip0.5em\hbox{for all}\hskip0.5em
  i=1,\ldots,\nv,
  \enum
}
where $\sigma$ is a permutation of $1,\ldots,\nv$.
At the same time, all vertex indices $i$
in the line list
must be replaced by $\sigma^{-1}(i)$.

\vskip1ex
\item{$(\sT{2})$} The lines in the line list $\lnl$ may be reordered in
an arbitrary way. In this case, the vertex list is not touched and
the line list is replaced by a new list $\lnl'$ with elements
\equation{
  \lnl'[j]=\lnl[\tau(j)]\hskip0.5em\hbox{for all}\hskip0.5em
  j=1,\ldots,\nl,
  \enum
}
where $\tau$ is a permutation of $1,\ldots,\nl$.
In addition, some of the boson lines
may have their ends interchanged, i.e.~the corresponding
line items are changed according to
\equation{
  ({\rm b},i_1,i_2)\to({\rm b},i_2,i_1).
  \enum
}
The total operation then consists of a permutation of the lines
and an interchange of the ends of some boson lines.

\parindent=1em
\vskip1ex\noindent
Diagram data $(\vxl,\lnl)$ and $(\vxl',\lnl')$
that can be matched by applying a product of the transformations
$\sT{1}$ and $\sT{2}$
are considered to be equivalent. By construction, there is a one-to-one
correspondence between diagrams and
equivalence classes of diagram data.

\subsection 4.3 Symmetry factors

The transformations $\sT{1}$ and $\sT{2}$ of the diagram data generate a group
of transformations with
$\nv!\,\nl!\,2^{\nb}$ elements, where $\nb$ denotes the number of
boson lines. Some of these transformations preserve the diagram
data, i.e.~are such that the vertex and line lists remain unchanged.
The number of elements of this subgroup of
transformations is the same for all
equivalent choices of the diagram data
and is referred to as the {\it symmetry factor of the diagram}.

It is understood here that
the field indices at the ends of the external and isolated lines
are distinguished by their names. As a consequence,
the transformations that preserve the diagram data cannot involve the
external and isolated lines. The vertices where
the external lines are attached are then also left untouched,
since these are distinguished by the field indices at the ends of
the attached external lines.

Given the diagram data $(\vxl,\lnl)$,
the symmetry factor of the diagram may therefore be calculated as the product
of the following factors:

\parindent=2.75em
\vskip1ex
\item{$(\sF{1})$}
The number of transformations $\sT{1}$ that permute
the internal vertices (those without
attached external lines) such that the
vertex list remains the same, while the
associated change of the line list can be undone by
a transformation $\sT{2}$.

\vskip1ex
\item{$(\sF{2})$}
A factor $n!$ for each instance, where two vertices
are connected by $n$ (and not more than $n$) boson lines or
where $n$ boson lines go from a vertex to itself.

\vskip1ex
\item{$(\sF{3})$}
A factor $2$ for each boson line that goes from a vertex
to itself.

\parindent=1em
\vskip1ex\noindent
The product of the factors $\sF{2}$ and $\sF{3}$
actually coincides with the number
of transformations $\sT{2}$ that preserve the line list.

The factors $\sF{1}$--$\sF{3}$ do not refer to the labeling of the external
lines. A symmetry factor associated with the latter is
\parindent=2.75em
\vskip1ex

\item{$(\sF{4})$} The number of permutations of the field indices
attached to the external lines, which leave the diagram data
unchanged modulo transformations $\sT{1}$ and $\sT{2}$.

\parindent=1em
\vskip1ex\noindent
The diagrams contributing to the correlation function (3.6), for example,
divide into sets of $n!(m!)^2/\sF{4}$ diagrams that coincide with
one another except for a permutation of the field indices
$a_1,\ldots,\beta_m$.

\section 5. Tree diagrams

The trivializing maps constructed in the following sections
are given by labeled tree diagrams like the ones shown in fig.~2.
In the general theory considered here, many more
diagrams can be drawn and there are both boson and fermion
random fields.
The random fields are introduced in this section
and various classes of tree diagrams are then defined.

\subsection 5.1 Random fields

Let $\eta_a$ be a real-valued Gaussian random field with
boson-field index $a$,
mean zero and variance
\equation{
  \langle\eta_a\eta_b\rangle=\delta_{ab}.
  \enum
}
Since the boson propagator $D^B_{ab}$ is a real symmetric matrix
with positive eigenvalues, it it can be written as the square of
another real symmetric matrix $M_{ab}$.
The field
\equation{
  \chi_a=M_{ab}\eta_b
  \enum
}
is then a Gaussian random field with variance
\equation{
  \langle\chi_a\chi_b\rangle=D^B_{ab}
  \enum
}
and mean zero.

Random fermion and antifermion fields are treated differently
and are given by
\equation{
  \zeta_{\alpha}=D^F_{\alpha\beta}\rho_{\beta},
  \qquad
  \zetabar_{\beta}=\rho_{\beta}^{\ast},
  \enum
}
where $\rho_{\alpha}$ is
a complex Gaussian random field
with fermion-field index $\alpha$, mean zero and variance
\equation{
  \langle\rho_{\alpha}\rho_{\beta}\rangle=0,\qquad
  \langle\rho_{\alpha}\rho_{\beta}^{\ast}\rangle=\delta_{\alpha\beta}.
  \enum
}
The two-point function
\equation{
  \langle\zeta_{\alpha}\zetabar_{\beta}\rangle=D^F_{\alpha\beta}
  \enum
}
then reproduces the fermion propagator, but the fact
that random fermion fields are pseudo-fermion fields (i.e.~complex
fields with fermion-field indices) should be kept in mind.

Further
random boson fields $\chi_{k,a}$, $k=1,2,\ldots$,
and random fermion fields $\zeta_{k,\alpha}$
(together with the antifermion
fields $\zetabar_{k,\alpha}$) will be needed as well.
The fields in these sequences are referred to
as the {\it labeled random fields}.
All random fields are assumed to be statistically independent
from one another.

Random fields can be attached to the vertices of a Feynman diagram
by replacing external lines through random fields of the appropriate
type (see fig.~5).
The value of such a diagram depends on the random fields and
is therefore a stochastic observable. It
is calculated by substituting the expressions for the propagators,
vertices and random fields for the corresponding graphical elements
and by contracting the field indices at the vertices.
There is no intervening propagator between the random fields and
the vertices to which they are attached.

\topinsert
\vbox{
\vskip0ex

\epsfxsize=6.5cm\hskip2.75cm\epsfbox{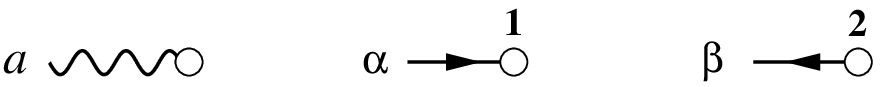}\hfill

\vskip2.0ex
\figurecaption{%
Random fields attached to the vertices of a diagram are represented
by external lines ending in an open circle.
From left to right, the random fields represented by the lines
in this figure are
$\chi_a$, $\zeta_{1,\alpha}$ and $\zetabar_{2,\beta}$.
The labels drawn near the circles are the indices of the
labeled random fields.
}
}
\vskip0ex
\endinsert

\subsection 5.2 Types of tree diagrams

A tree diagram is a connected diagram with one or more
vertices and no loops. The following
types of tree diagrams with attached random fields will be considered:

\parindent=2.0em
\vskip1ex
\item{(1)}{\it Unlabeled diagrams.}
All external lines of these tree diagrams represent one of the random fields
$\chi_{a}$, $\zeta_{\alpha}$
or $\zetabar_{\beta}$ and are therefore unlabeled.

\vskip1ex
\item{(2)}{\it Labeled diagrams.}
These diagrams are obtained
by replacing some of the random fields at
the leaves of an unlabeled tree diagram
through labeled random fields.
The latter must come in pairs
with matching type and index.
Moreover, each pair may occur at most once and
the ends of each fermion path in the diagram
must be assigned the same label or remain
both unlabeled.

\vskip1ex
\item{(3)}{\it Rooted labeled diagrams.}
The diagrams of this kind are labeled tree diagrams, where
one unlabeled external boson line is replaced by an
ordinary external boson line (i.e.~by a line representing
the boson propagator).

\parindent=1em
\vskip1ex\noindent
Labeled diagrams need not have any labeled external lines,
i.e.~the unlabeled tree diagrams are included in the set of labeled
diagrams. An example of each type of diagram is shown in fig.~6.

\topinsert
\vbox{
\vskip0ex

\epsfxsize=10.4cm\hskip0.9cm\epsfbox{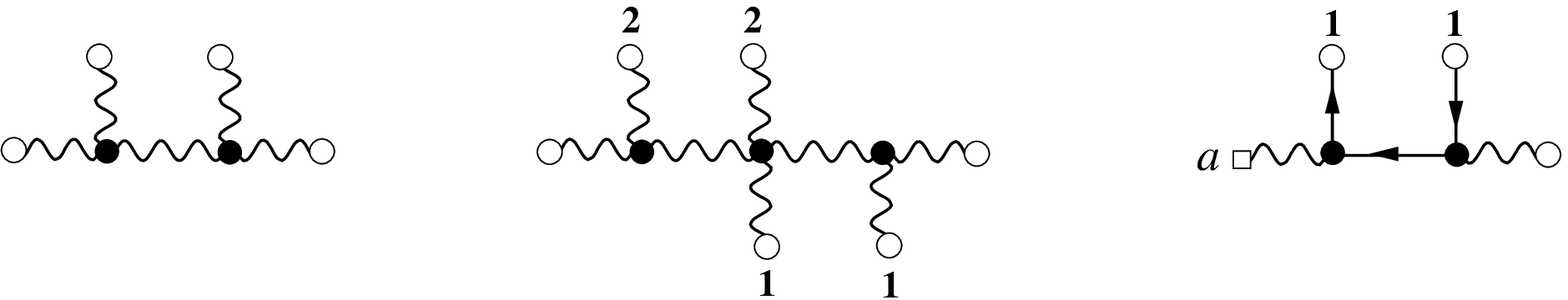}\hfill

\vskip1.0ex
\figurecaption{%
Examples of unlabeled, labeled and rooted labeled tree diagrams.
Random-field
labels must come in pairs, but boson and fermion labels may be the same,
since they refer to different sequences of random fields. For simplicity,
the vertex labels are omitted in this figure.
}
}
\vskip0ex
\endinsert

\section 6. Structure of the trivializing map

The strategy in the following is to write down
a structured ansatz for the trivializing stochastic field
and to show that the trivializing property, eq.~(2.3),
is guaranteed, if the adjustable parameters are chosen
appropriately.

\subsection 6.1 Diagram sets

The construction starts by introducing certain sets
\equation{
  \uset_{k,l},\quad k=1,2,\ldots,\quad l=0,1,\ldots,\lbar(k),
  \enum
}
of labeled tree diagrams, where $k$
is the order in the coupling
and $l$ the number of labeled pairs of external lines
of the diagrams in $\uset_{k,l}$.
Since tree diagrams of order $k$ have
at most $k+2$ external lines, the maximal number $\lbar(k)$ of labeled pairs
cannot be larger than $\lfloor k/2\rfloor+1$.

$\uset_{k,0}$ coincides with the set of unlabeled
tree diagrams of order $k$.
For $l\geq1$ the sets $\uset_{k,l}$ are
obtained by going through the diagrams in $\uset_{k,0}$
and labeling $l$ pairs of external lines
in all possible ways. For each type of field (boson or fermion),
the pairs are labeled from 1 in steps of 1.
Diagrams that coincide after contracting
the labeled external lines
are not distinguished and only one of them
is to be included in $\uset_{k,l}$.

Another family of diagram sets, $\rset_{k,l}$, derives from
the sets $\uset_{k,l}$ by selecting the diagrams
with at least one unlabeled external boson line and no
unlabeled external fermion lines.
Each of these diagrams is then transformed to a rooted labeled tree diagram
by replacing an unlabeled external boson line through an ordinary
external boson line. For notational convenience, the set $\rset_{0,0}$,
whose only element is the first diagram in fig.~5, is included in this
family of sets.

The construction of the sets $\rset_{k,l}$ of diagrams depends
on some arbitrary choices to be made
in the step leading from $\uset_{k,0}$ to $\uset_{k,l}$ and
when the root line is selected.
There are, in principle, alternative constructions, where
the diagrams in $\rset_{k,l}$ are replaced
by linear combinations diagrams,
but the number of diagrams contributing to the trivializing
field is then likely to increase.

\subsection 6.2 Ansatz for the trivializing map

As in numerical lattice QCD, the fermion fields
$\psi_{\alpha}$ and $\psibar_{\beta}$
are integrated out and the primary correlation functions considered
are those of the boson field $\varphi_a$.
The trivializing stochastic field $\tmap_a$
is then a boson field too, but
must take the effects of the fermion loops
into account so that the correlation functions of the full
theory are properly reproduced.

Following the lines of sect.~2,
the field $\tmap_a$ is expanded in a series
\equation{
  \tmap_a=\sum_{k=0}^{\infty}\tmap_{k,a},
  \enum
  \nexteq{2.0ex}
  \tmap_{k,a}=\sum_{l=0}^{\lbar(k)}\sum_{\dgmr\in\rset_{k,l}}
  c(\dgmr)v(\dgmr)_a,
  \enum
}
where $v(\dgmr)_a$ denotes the value of the diagram $\dgmr$
with field index $a$ at the end of the root line.
The coefficient $c(\dgmr)$ of the leading term,
\equation{
  \tmap_{0,a}=\chi_a,
  \enum
}
is set to unity, but the coefficients of the tree diagrams
are left unspecified at this point.
With increasing order in the coupling,
the trivializing field depends on more and more random fields.
Their number however grows only linearly
with the maximal number of loops Feynman diagrams
can have at the order considered.

The ansatz (6.3) and the way in which the fermion random fields
were introduced in general leads to complex trivializing fields.
In purely bosonic theories, the field is real if (and only if)
the action has no imaginary part.
Complex trivializing fields do not give rise to any conceptual difficulties
in perturbation theory, since the expectation values
of all observables are eventually expressed through the basic correlation
functions of the boson field and these are correctly reproduced
by the trivializing field.

\section 7. Determination of the coefficients $\mib c(\dgmr)$

The coefficients of the tree diagrams in the expansion (6.3)
of the trivializing stochastic field are to be chosen such that
\equation{
  \langle\tmap_{a_1}\ldots\tmap_{a_n}\rangle=
  \langle\varphi_{a_1}\ldots\varphi_{a_n}\rangle
  \enum
}
for all $n\geq1$ and to all orders in the coupling.
At any given order, both correlation functions in eq.~(7.1)
evaluate to sums of ordinary diagrams,
the ones on the left being obtained by Wick-contracting the
random fields on which
the stochastic fields $\tmap_{a_1},\ldots,\tmap_{a_n}$ depend.
The equation thus holds algebraically if
the diagrams and their coefficients
are the same on the left and the right.

\subsection 7.1 Disconnected parts

The connected and disconnected parts of the correlation functions are
related to each other through the moment-cumulant transformation
and thus in a way that does not refer to any particular properties
of the correlation functions.
Equation (7.1) is therefore satisfied if and only if
the connected parts of the correlation functions are the same.

Since tree diagrams are
connected, the connected parts
of the correlation functions of the stochastic field
coincide with the sum of the connected diagrams
obtained by the Wick contraction of the random fields.
As a consequence, it suffices to ensure that
the connected diagrams and their coefficients
are the same on the two sides of eq.~(7.1)
for the equation to be fully satisfied.

\subsection 7.2 Bose symmetry

The connected diagrams contributing to the correlation function
$\langle\varphi_{a_1}\ldots\varphi_{a_n}\rangle$ divide into subsets of
diagrams that coincide with one another up to a permutation of
the field indices $a_1,\ldots,a_n$.
In general, there are fewer than
$n!$ diagrams in a subset, but it always contains all diagrams
with different index assignments.
Since the diagrams have the same symmetry and sign factors,
the sum of their contributions
is totally symmetric in the field indices and
given by
\equation{
  {(-1)^{\nlp}\over\sF{1}\sF{2}\sF{3}\sF{4}}
  \sum_{\pi\in\Sn}v(\dgm)_{a_{\pi(1)}\ldots a_{\pi(n)}},
  \enum
}
where $\dgm$ is any diagram in the subset, $\nlp$ the number
of fermion loops in the diagram and
$\Sn$ the group of permutations of $1,\ldots,n$.
The product of the symmetry factors
$\sF{1}\ldots\sF{4}$ includes $\sF{4}$ in
order to correct for the fact that not all index permutations
yield different diagrams (cf.~subsect.~4.3).

Practically the same remarks apply in the case of
the diagrams generated by the Wick contractions
on the left of eq.~(7.1). Since the
stochastic field is bosonic, its correlation functions
may be written in the manifestly symmetric form
\equation{
  \langle\tmap_{a_1}\ldots\tmap_{a_n}\rangle=
  \noenum
  \nexteq{2.0ex}
  \qquad
  \sum_{\dgmr_1\in\rset}\ldots\sum_{\dgmr_n\in\rset}
  c(\dgmr_1)\ldots c(\dgmr_n)
  {1\over n!}\sum_{\pi\in\Sn}
  \langle v(\dgmr_1)_{a_{\pi(1)}}\ldots v(\dgmr_n)_{a_{\pi(n)}}\rangle.
  \enum
}
The sums in this equation run over the union
$\rset$ of the sets $\rset_{k,l}$ of diagrams.
They are formal sums that only
make sense at any finite order in the coupling.

The contractions of the random fields on the right of eq.~(7.3)
produce terms proportional to
\equation{
  \sum_{\pi\in\Sn}v(\dgm)_{a_{\pi(1)}\ldots a_{\pi(n)}},
  \enum
}
where $\dgm$ is a Feynman diagram with $n$ ordinary boson external lines
(and no further external lines).
Matching the two sides of eq.~(7.1) thus amounts to matching the
coefficients of these index-symmetrized diagram values.

\subsection 7.3 Are there more equations than unknowns?

The discussion in the previous subsections shows that there is
one coefficient equation
per connected diagram $\dgm$ with $n\geq1$
ordinary external boson lines,
where diagrams related by a permutation of the
field indices $a_1,\ldots,a_n$ labeling the external
lines are not distinguished.
In the following, the notation $[\dgm]$ is used for
diagrams $\dgm$ modulo field-index permutations.

The ansatz for the trivializing stochastic field, on the
other hand, depends on one adjustable coefficient $c(\dgmr)$
per rooted labeled tree diagram $\dgmr\in\rset_{k,l}$.
By construction, each diagram $\dgmr$ derives from
a unique labeled tree diagram in $\uset_{k,l}$,
which in turn maps to an ordinary Feynman diagram $[\dgm]$
when its labeled external lines are contracted and
the unlabeled ones are replaced by ordinary external lines.
This relation between $\dgmr$ and $[\dgm]$ is one-to-one and such that
\equation{
  \sum_{\pi\in\Sn}
  \langle v(\dgmr)_{a_{\pi(1)}}\tmap_{0,a_{\pi(2)}}\ldots
  \tmap_{0,a_{\pi(n)}}\rangle
  \noenum
  \nexteq{1.5ex}
  \qquad
  =(n-1)!\sum_{\pi\in\Sn}v(\dgm)_{a_{\pi(1)}\ldots a_{\pi(n)}}
  +\ldots,
  \enum
}
where the ellipsis stands for the contributions of the disconnected
diagrams produced by the contractions of the random fields
on the left of the equation.
Moreover, all diagrams $[\dgm]$ are obtained from a diagram $\dgmr$
in this way.

At any given order in the coupling,
there are therefore exactly as many equations as there
are unknown coefficients.

\subsection 7.4 Recursive solution of the equations

The equations for the coefficients $c(\dgmr)$
may now be solved one after the other
in a particular order.
As explained in the previous subsections, there is
one coefficient equation per connected ordinary diagram $[\dgm]$
with right-hand side equal to
\equation{
   {(-1)^{\nlp}\over\sF{1}\sF{2}\sF{3}\sF{4}}.
   \enum
}
On the left of the equation, there is a sum of terms that are obtained by
evaluating the expectation values in eq.~(7.3).
More precisely, the term contributed by
a given combination $\dgmr_1,\ldots,\dgmr_n$ of rooted labeled tree
diagrams is given by
\equation{
   c(\dgmr_1)\ldots c(\dgmr_n){N_{\rm c}\over n!},
   \enum
}
with $N_{\rm c}$ being the number of Wick contractions
of the random fields in the product
$v(\dgmr_1)_{a_1}\ldots v(\dgmr_n)_{a_{n}}$
which yield the diagram $[\dgm]$.

There are two types of diagram combinations $\dgmr_1,\ldots,\dgmr_n$
that can contract to the diagram $[\dgm]$:

\parindent=2.0em
\vskip1ex
\item{(1)} Two or more of the diagrams are tree diagrams and are thus
contained in some of the sets $R_{k,l}$ with $k\geq1$.
The order in the coupling of all diagrams $\dgmr_1,\ldots,\dgmr_n$
must then be strictly less than the one of $\dgm$.

\vskip1ex
\item{(2)} Only one of the diagrams is non-trivial,
say $\dgmr_1\in\rset_{k,l}$ for some $k\geq1$ and $l\geq0$.
The order $k$ of the diagram
must then be the same as the one of $\dgm$. Moreover, the number
$l$ of labeled pairs of external lines
cannot exceed the number of loops of $\dgm$.
If $l$ coincides with the number of loops,
the construction of the sets $\rset_{k,l}$ of diagrams implies
that $\dgmr_1$ must be equal to $\dgmr$, the rooted labeled
tree diagram associated with $[\dgm]$.

\parindent=1em
\vskip1ex\noindent
The equations for the coefficients can therefore be solved recursively
with increasing order $k$ in the coupling and, at fixed $k$, increasing
loop number $l$. For a given diagram $[\dgm]$,
the equation is of the form
\equation{
  c(\dgmr)+\hbox{already known terms}=
  {(-1)^{\nlp}\over\sF{1}\sF{2}\sF{3}\sF{4}}
  \enum
}
and can therefore be solved for the coefficient $c(\dgmr)$
of the tree diagram $\dgmr$ associated with $[\dgm]$.

\subsection 7.5 Automated computation of the coefficients

The number of tree diagrams contributing to the trivializing
stochastic field at order $k$ increases roughly like $k!$.
An automated computation of their coefficients is therefore
desirable and can follow essentially the steps described in
this section.

First the various sets of diagrams must be generated. It is advisable
to introduce a partial ordering of the diagrams based on
their structure and to build up binary search trees [\ref{KnuthIII}]
together with the sets. The computer time spent to find the
diagrams generated by the contractions of the random fields
on the right of eq.~(7.3) otherwise grows with the square
of the number of diagrams.

The diagram ordering also allows the number of terms
contributing to the
sum over diagrams in eq.~(7.3) to be reduced by a large factor
by summing over
ordered combinations $\dgmr_1,\ldots,\dgmr_n$ of diagrams only.
Another acceleration of the calculations can be achieved
by collecting contractions of random fields, which
only differ by a permutation of some vertex legs of equal type
and therefore obviously yield the same contracted diagram.

\section 8. Trivializing map in lattice QCD

The abstract theory considered in the previous sections
includes all popular formulations of lattice QCD as special cases,
provided the chosen boundary conditions in finite volume are such
that the gauge modes are the only zero modes of the action.
This property is guaranteed with Schr\"odinger-functional (SF)
[\ref{SF},\ref{SFquark}]
and open-SF boundary conditions [\ref{openSF}], for example.
The aim in the present section is to describe and further develop
the trivializing map in lattice QCD, using a slightly less abstract
language.

\subsection 8.1 Fields, action and parameters

In perturbation theory, the gauge-field variables $U(x,\mu)\in\SUthree$
may be represented through a real-valued
gauge potential $A^a_{\mu}(x)$ according to
\equation{
   U(x,\mu)=\exp\{g_0A^a_{\mu}(x)T^a\},
   \enum
}
where $g_0$ denotes the bare gauge coupling and $T^a$, $a=1,\ldots,8$,
a basis of anti-Hermitian $\SUthree$ generators
(for notational convenience, the lattice spacing is set to unity).
The gauge degrees of freedom then need to be fixed by adding
a gauge-fixing term and the associated Faddeev--Popov ghosts
to the theory. Usually a lattice version of
the covariant gauge with bare gauge parameter $\lambda_0$
is chosen.

Further parameters of the lattice theory are
the number $\Nf$ of quark flavours, the quark masses
and optionally
a set of irrelevant parameters, the so-called improvement
coefficients, which are tuned so as
to reduce the effects of the non-zero lattice
spacing on the calculated physical quantities.
For simplicity,
the dependence of the improvement coefficients
on the gauge coupling is however ignored in the following and
the quarks are assumed to be mass-degenerate with bare mass
parameter $m_0$.

\subsection 8.2 Bare and renormalized perturbation expansion

The bare perturbation expansion of the correlation functions
of the gauge potential is obtained by
inserting the representation (8.1) of the link variables
in the action and by
expanding the functional integral in powers of $g_0$.
In addition to the gluon vertices that derive
from the gauge action, the expansion of
the integration measure gives rise to further
vertices of order $g_0^l$ with $l=2,4,\ldots$ gluon legs.
There are quark-gluon vertices with any number
of gluon legs and ghost-gluon vertices with $1,2,4,\ldots$
gluon legs
(see ref.~[\ref{SelectedTopics}], for example, for all-order formulae
for the measure and ghost-gluon vertices).

The renormalization of the gauge coupling and the gauge potential
merely amounts to a reordering of the bare perturbation expansion,
but the renormalization of the gauge-fixing parameter and the quark
mass requires additional
diagrams with gauge-fixing and mass vertex insertions
to be calculated.
This complication can economically be dealt with by substituting
\equation{
  \lambda_0=\lambda+\delta\lambda,
  \qquad
  m_0=m+\delta m,
  \enum
}
for the bare gauge-fixing and mass parameters,
where $\lambda$ and $m$ are held fixed, while the terms
proportional to $\delta\lambda$ and $\delta m$ in the action
are treated as perturbations of order $g_0^2$.
The exact form of the associated vertices depends on the chosen
lattice action, but they normally have only two legs.

\subsection 8.3 Trivializing map to second order in
                the coupling

The rooted labeled tree diagrams contributing to the
stochastic gauge potential that trivializes the lattice theory
to a given order of perturbation theory
are made of the vertices mentioned in subsect.~8.2
and the gluon, quark and ghost propagators.
In order to minimize the number of diagrams
and still be able to easily pass to the renormalized
expansion, the best strategy is to expand
the field in the bare coupling and to include
the gauge-fixing and quark-mass vertices in the Feynman rules.

Random quark fields must normally carry a flavour index,
but for mass-degenerate quarks,
trivializing gauge potentials can be constructed
with flavour-less random fields and coefficients
$c(\dgmr)$ that depend on $\Nf$.
The first- and second-order terms in the expansion of the
trivializing field are then given by
\equation{
  {\cal A}_1=\frac{1}{6}v(\dgmr_1),
  \enum
  \nexteq{2.5ex}
  {\cal A}_2=\frac{1}{2}v(\dgmr_2)
            +\frac{1}{2}v(\dgmr_3)
            +\frac{1}{24}v(\dgmr_4)
            +\frac{1}{8}v(\dgmr_5)
            -\frac{1}{2}v(\dgmr_6)
            -\frac{1}{2}\Nf v(\dgmr_7)
  \noenum
  \nexteq{1.5ex}
  {\phantom{{\cal A}_2={}}}
            +\frac{5}{72}v(\dgmr_8)
            +\frac{1}{12}v(\dgmr_9)
            -\frac{1}{2}v(\dgmr_{10})
            -\frac{1}{2}\Nf v(\dgmr_{11}),
  \enum
}
where, for brevity,
the coordinates and indices $x,\mu,a$ have been suppressed. The
diagrams $\dgmr_1,\ldots,\dgmr_{11}$ are shown in fig.~7.
There are no diagrams $\dgmr$ in this list,
where the associated contracted diagram $[\dgm]$ has
tadpole subdiagrams%
\kern1pt\footnote{$\dagger$}{\footnotefont%
A tadpole subdiagram is a subdiagram without external lines, which
is connected to the rest of the diagram through a single gluon line.},
since $v(\dgm)_{a_1\ldots a_n}=0$ in this case and the associated coefficient
equation consequently does not need to be satisfied.
The coefficient $c(\dgmr)$ may therefore be set to zero
(cf.~sect.~7).

\topinsert
\vbox{
\vskip0ex

\epsfxsize=13.2 true cm\hskip-0.0 true cm\epsfbox{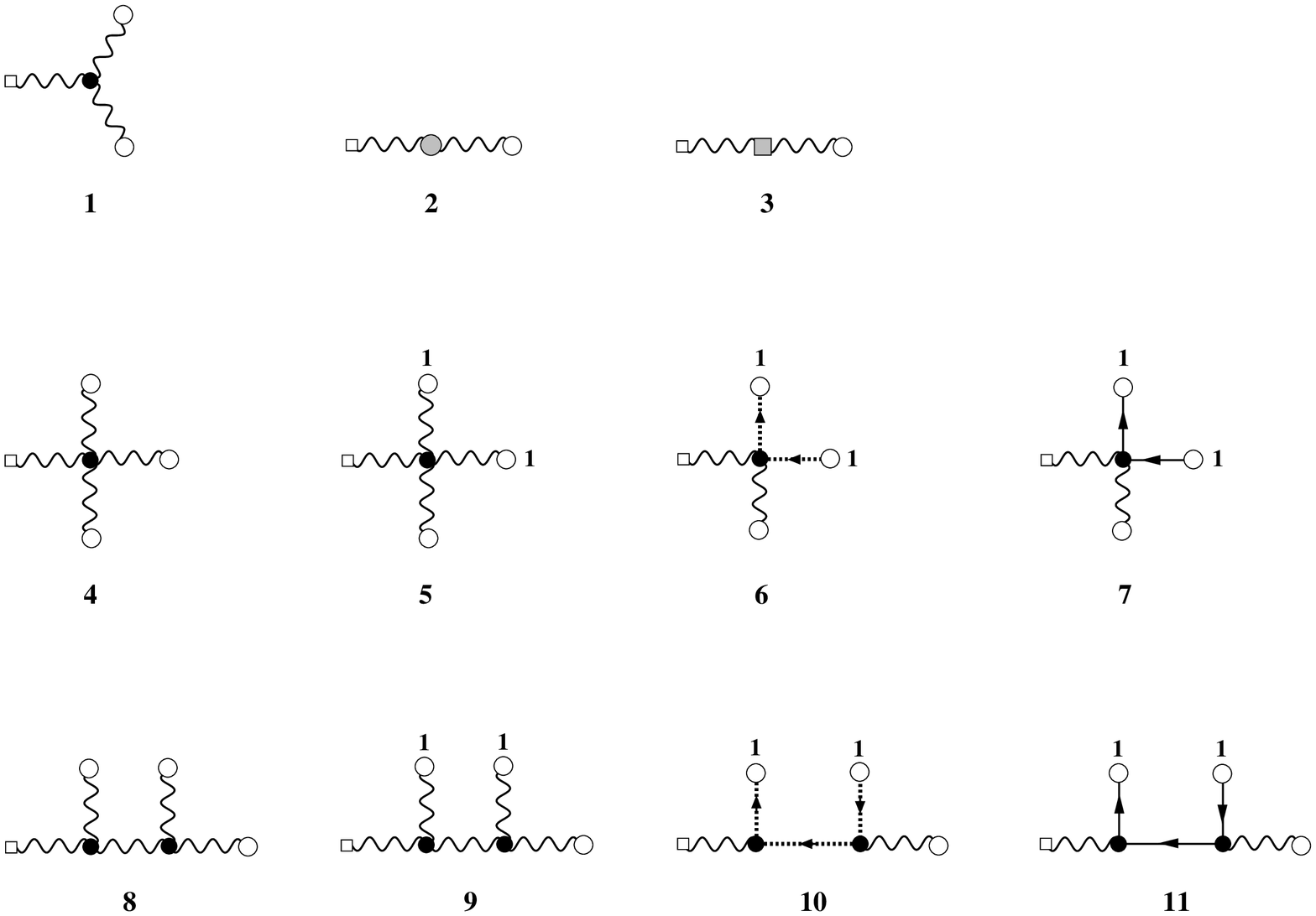}\hfill

\vskip3.0ex
\figurecaption{%
Rooted labeled tree diagrams $\dgmr_1,\ldots,\dgmr_{11}$
contributing to the trivializing
gauge potential at first
order in the gauge coupling (diagram $1$)
and at second order (diagrams $2-11$).
Shaded square vertex symbols represent
the vertices that derive from the gauge-field integration measure
and shaded circular symbols stand for the insertion of a
gauge-fixing vertex proportional to $\delta\lambda$.
Quark and ghost lines (solid and dotted)
are distinguished in these diagrams.
}
}
\vskip0ex
\endinsert

\subsection 8.4 Higher-order trivializing maps

Apart from the growing combinatorial complexity,
the construction of the trivializing gauge potential
at higher orders in the coupling
does not run into additional difficulties.
Quark-mass vertices and
tree diagrams with more than one
branch first appear at fourth order.
At order $k$, the coefficients of the tree diagrams
are polynomials in $\Nf$
of degree up to $\lfloor k/2\rfloor$
with rational coefficients.

In view of the large number of diagrams,
the generation of the diagram sets $\rset_{k,l}$ and
the calculation of the coefficients of the diagrams
rapidly becomes a task that calls for automation.
As explained in subsect.~8.3, the tree diagrams that contract
to diagrams with tadpole subdiagrams can be omitted from the
beginning. The other tree diagrams can be generated
following the lines of subsect.~6.1.
Some arbitrary choices need to be made at this point,
but the numbers of diagrams quoted in
the second and forth column of table~1 do not depend on these.

Even with a straightforward
selection rule like the one used to produce the figures
in table~1, surprisingly many diagrams
turn out to have vanishing coefficient
(see 3rd column of table~1).
If the diagrams with gauge-fixing and ghost vertices are omitted,
the numbers of diagrams are reduced by about a factor $2$
(columns 4 and 5).
These diagrams can actually be dropped in the case of
open-SF boundary conditions [\ref{openSF}], for example,
if the temporal gauge
$A^a_0(x)=0$ is chosen.

\topinsert
\newdimen\digitwidth
\setbox0=\hbox{\rm 0}
\digitwidth=\wd0
\catcode`@=\active
\def@{\kern\digitwidth}
\tablecaption{Numbers of rooted labeled tree diagrams $\dgmr$ in lattice QCD}
\vskip-3.0ex

$$\vbox{\settabs\+&%
                  xxxxxxx&xx&%
                  xxxxxxxxxxxx&xx&%
                  xxxxxxxxxxxx&xx&%
                  xxxxxxxxxxxx&xx&%
                  xxxxxxxxxxxx&xx\cr
\thicktablerule
\vskip1.2ex
                \+& \hfill order\hfill
                 && \hfill all\hfill
                 && \hfill $c(\dgmr)\neq0$\hfill
                 && \hfill no ghosts\hfill
                 && \hfill $c(\dgmr)\neq0$\hfill
                 &\cr
\vskip0.8ex
\thintablerule
\vskip1.2ex
  \+& \hfill $1$\hfill
  &&  \hfill $1$\hskip1.5em
  &&  \hfill $1$\hskip1.5em
  &&  \hfill $1$\hskip1.5em
  &&  \hfill $1$\hskip1.5em
  &\cr
  \+& \hfill $2$\hfill
  &&  \hfill $10$\hskip1.5em
  &&  \hfill $10$\hskip1.5em
  &&  \hfill $7$\hskip1.5em
  &&  \hfill $7$\hskip1.5em
  &\cr
  \+& \hfill $3$\hfill
  &&  \hfill $19$\hskip1.5em
  &&  \hfill $19$\hskip1.5em
  &&  \hfill $14$\hskip1.5em
  &&  \hfill $14$\hskip1.5em
  &\cr
  \+& \hfill $4$\hfill
  &&  \hfill $146$\hskip1.5em
  &&  \hfill $141$\hskip1.5em
  &&  \hfill $91$\hskip1.5em
  &&  \hfill $87$\hskip1.5em
  &\cr
  \+& \hfill $5$\hfill
  &&  \hfill $522$\hskip1.5em
  &&  \hfill $489$\hskip1.5em
  &&  \hfill $309$\hskip1.5em
  &&  \hfill $286$\hskip1.5em
  &\cr
  \+& \hfill $6$\hfill
  &&  \hfill $4042$\hskip1.5em
  &&  \hfill $3524$\hskip1.5em
  &&  \hfill $2106$\hskip1.5em
  &&  \hfill $1816$\hskip1.5em
  &\cr
  \+& \hfill $7$\hfill
  &&  \hfill $20312$\hskip1.5em
  &&  \hfill $16851$\hskip1.5em
  &&  \hfill $9978$\hskip1.5em
  &&  \hfill $8274$\hskip1.5em
  &\cr
  \+& \hfill $8$\hfill
  &&  \hfill $159190$\hskip1.5em
  &&  \hfill $127143$\hskip1.5em
  &&  \hfill $70854$\hskip1.5em
  &&  \hfill $56986$\hskip1.5em
  &\cr
\vskip1.2ex
\thicktablerule
}
$$
\vskip0.0ex
\endinsert

\section 9. Concluding remarks

The viability of the trivializing fields
constructed in this paper for numerical stochastic
perturbation theory still needs to be demonstrated.
As already mentioned,
rooted labeled tree diagrams can be evaluated
through a recursive procedure that starts
from the leaves of the diagrams and visits the vertices
one after another until the root line is reached.
At each vertex the attached fields and branches are combined
to form a new field, which is then propagated to the
next vertex. Highly efficient algorithms
(the fast Fourier transform [\ref{FFT}]
and multigrid methods [\ref{Saad}])
are available for the field propagation, which amounts
to solving a free-field equation with known right-hand side.

The rapid growth of the number of diagrams with the order in the
coupling potentially sets a sharp limit on the applicability
of instantaneous stochastic perturbation theory.
There are, however, various ways to organize the computations
efficiently, exploiting the product structure of tree diagrams.
Reusing already calculated branches, for example,
or factoring out the vertices recursively, starting from the root line,
may result in significant speed-up factors and thus
allow the calculations to be driven to higher
orders than appears to be practically feasible at first sight.

An intriguing aspect of instantaneous stochastic perturbation theory
is the fact that theories with complex action
are not excluded and do not even require a special treatment.
In particular, chiral gauge theories [\ref{ChGauge}]
and QCD at finite baryon density are among the accessible theories.

\beginbibliography


\bibitem{SQI}
G. Parisi, Y.-S. Wu,
{\it Perturbation theory without gauge fixing},
Sci. Sin. 24 (1981) 483.

\bibitem{SQII}
P. H. Damgaard, H. H\"uffel,
{\it Stochastic quantization},
Phys. Rept. 152 (1987) 227.


\bibitem{SPThI}
F. Di Renzo, G. Marchesini, P. Marenzoni, E. Onofri,
{\it Lattice perturbation theory on the computer},
Nucl. Phys. Proc. Suppl. 34, 795 (1994).

\bibitem{SPThII}
F. Di Renzo, E. Onofri, G. Marchesini, P. Marenzoni,
{\it Four loop result in SU(3) lattice gauge theory by a stochastic method:
Lattice correction to the condensate},
Nucl. Phys. B426, 675 (1994).


\bibitem{DiRenzoScorzato}
F. Di Renzo, L. Scorzato,
{\it Numerical stochastic perturbation theory for full QCD},
JHEP 10 (2004) 073.


\bibitem{BrambillaBridaEtAl}
M. Brambilla, M. Dalla Brida, F. Di Renzo, D. Hesse, S. Sint,
{\it Numerical stochastic perturbation theory in the Schr\"odinger functional},
PoS (Lattice 2013) 325.

\bibitem{BridaHesse}
M. Dalla Brida, D. Hesse,
{\it Numerical stochastic perturbation theory and the gradient flow},
PoS (Lattice 2013) 326.


\bibitem{LvinI}
G. G. Batrouni, G. R. Katz, A. S. Kronfeld, G. P. Lepage, B. Svetitsky,
K. G. Wilson,
{\it Langevin simulations of lattice field theories},
Phys. Rev. D32 (1985) 2736.

\bibitem{LvinII}
A. Ukawa, M. Fukugita,
{\it Langevin simulation including dynamical quark loops},
Phys. Rev. Lett. 55 (1985) 1854.

\bibitem{LvinIII}
G. S. Bali, C. Bauer, A. Pineda, C. Torrero,
{\it Perturbative expansion of the energy of static sources at large orders
in four-dimensional SU(3) gauge theory},
Phys. Rev. D87 (2013) 094517.


\bibitem{LvinIV}
A. S. Kronfeld,
{\it Dynamics of Langevin simulations},
Prog. Theor. Phys. Suppl. 111 (1993) 293.


\bibitem{TrivMap}
M. L\"uscher,
{\it Trivializing maps, the Wilson flow and the HMC algorithm},
Commun. Math. Phys. 293 (2010) 899.


\bibitem{KnuthIII}
D. E. Knuth,
{\it The art of computer programming}, Vol. 3
(Addison-Wesley, Reading, 1973).


\bibitem{SF}
M. L\"uscher, R. Narayanan, P. Weisz, U. Wolff,
{\it The Schr\"odinger functional ---
a renormalizable probe for non-Abelian gauge theories},
Nucl. Phys. B384 (1992) 168.

\bibitem{SFquark}
S. Sint,
{\it On the Schr\"odinger functional in QCD},
Nucl. Phys. B421 (1994) 135.


\bibitem{openSF}
M. L\"uscher,
{\it Step scaling and the Yang-Mills gradient flow},
JHEP 06 (2014) 105.


\bibitem{SelectedTopics}
M. L\"uscher,
{\it Selected topics in lattice field theory},
Lectures given at Les Houches (1988), in:
{\it Fields, strings and
critical phenomena}, eds. E. Br\'ezin and J. Zinn-Justin
(North-Holland, Amsterdam, 1989).


\bibitem{FFT}
C. van Loan,
{\it Computational frameworks for the fast Fourier transform},
Frontiers in applied mathematics, vol. 10 (SIAM, Philadelphia, 1992).


\bibitem{Saad}
Y. Saad,
{\it Iterative methods for sparse linear systems},
2nd ed. (SIAM, Philadelphia, 2003); see also
{\tt http://www-users.cs.umn.edu/\urltilde saad/}.


\bibitem{ChGauge}
M. L\"uscher,
{\it Lattice regularization of chiral gauge theories to all orders
of perturbation theory},
JHEP 06 (2000) 028.

\endbibliography

\bye